\documentclass[11pt]{article}
\usepackage{a4wide}
\usepackage{amsmath}
\usepackage{amsfonts}
\usepackage{amssymb}

\newcommand{\ba}{\begin{equation}}
\newcommand{\ea}{\end{equation}}
\usepackage{graphicx}

\def\one{\mathbf{1}}
\def\oneP#1{\one_{\tau}}

\newtheorem{prop}{Proposition}
\newtheorem{lemm}{Lemma}

\newtheorem{defi}{Definition}

\newtheorem{note}{Note}

\begin{document}
	
	\author{Cyril Grunspan\footnote{I wish to thank Yann Braouezec, Daniel Gabay, Peter Tankov, Joris Van der Hoeven.}\\
	ESILV,  Department of Financial Engineering\\
	92916 Paris La D\'efense Cedex \\
	cyril.grunspan@devinci.fr} 
	
	\title{ {Asymptotic Expansions of the Lognormal Implied Volatility : A Model Free Approach}}
	
	\maketitle
	
%	\begin{center}
%	{\it  Mathematical Finance ?} {\rm (Submitted - not yet! -)  }
%	\end{center} 
	
	\begin{abstract}
   \noindent We invert the Black-Scholes formula. We consider the cases low strike, large strike, short maturity and 
   large maturity. We give explicitly the first 5 terms of the expansions. A method to compute all the 
   terms by induction is also given. At the money, we have a closed form formula for implied lognormal volatility 
   in terms of a power series in call price.
	\end{abstract}

	\noindent JEL Classification G12 G13 C65 \bigskip
	

	\section{Introduction}
	\subsection{Overview}
	In a market with no arbitrage, the price of a call option can take two extreme values : 
	its ``intrinsic value'' which is equal to the payoff of the option (lower boundary value) and 
	the spot price (upper boundary value). For simplicity, we assume a market with no interest rate. 
	Otherwise, we would consider the forward price. We will consider here the case when the call price 
	is close to its boundary value and we will obtain in that case an approximation of the corresponding 
	lognormal implied volatility. This case happens in particular when the maturity of the option is small.
	To be precise, in the case when $T\rightarrow 0$ (resp. $T\rightarrow +\infty$), 
	we will obtain an asymptotic expansion of the implied 
	lognormal volatility as a sum of terms of the form $\lambda^{i}\ln^{j}(\lambda)$ with $j<i$ 
	and $\lambda=-\displaystyle\frac{1}{\ln\left(\frac{C(T,K)-(S-K)_+}{S}\right)}$
	(resp. $\lambda=-\displaystyle\frac{1}{\ln\left(\frac{S-C(T,K)}{S}\right)}$)
	where $C(T,K)$ denotes the price of a call option with strike $K$, maturity $T$ and spot price $S$
	(Proposition \ref{laprop}). Note that here, the spot price $S$ is present only to insure that the ratio 
	$\displaystyle\frac{C(T,K)-(S-K)_+}{S}$ (resp. $\displaystyle\frac{S-C(T,K)}{S}$)is with no-dimension. 
	The important quantity is the ``time-value'' ${\rm TV}(T,K):=C(T,K)-(S-K)_+$ 
	(resp. ``covered call'' ${\rm CC(T,K)}:=S-C(T,K)$)
	The computations involve no complicated formulas except may be a well known asymptotic expansion for the 
	incomplete Gamma function (Equation (\ref{wellknown})). The interest of such a 
	formula is twofold. First, it gives quickly an easy approximation of the true implied lognormal volatility. 
	This can serve as a starting point for the calculus of the exact implied lognormal volatility using a Newton 
	method for instance. The formula can also be useful to transform theoretical approximations of a call price into 
	approximations of implied lognormal volatility. Indeed, asymptotics of call prices can be obtained with the help of 
	stochastic differential equations of partial differential equations using perturbation methods. 
	Then, a transformation has to be made to obtain the implied lognormal volatility which is of a fundamental interest 
	for the practitioner.	
	
	All our work is based on a single inversion formula. Explicitly we invert the following equation for $\lambda\ll 1$
	and $\beta>0$ (see Note \ref{thenote}):
	\begin{equation}
	\label{tout}
	v^{\beta}{\rm e}^{-\frac{1}{v}}\,\left[\displaystyle\sum_{k=0}^{N}\, \alpha_k\, 
	v^k\,+O\left(v^{N+1}\right)\right] ={\rm e}^{\gamma}\,{\rm e}^{-\frac{1}{\lambda}}  
	\end{equation}
	
	\noindent
	The good framework for solving this problem (i.e. obtain $v$ in terms of $\lambda$) is the theory of transseries 
	(see \cite{VdH}). In the expansion of $v$ in terms of $\lambda$ coming from (\ref{tout}) it is important to go up to order
	$5$ (for us, order $0$ is $\lambda$, order $1$ is $\lambda^2\ln(\lambda),\,...$ order $5$ is $\lambda^3$)
	to see $\alpha_1$ (see Lemma \ref{da}):
	\begin{equation}
	\nonumber
	v=\lambda-\beta\lambda^2
	\ln\lambda+\gamma\lambda^2+\beta^2\lambda^3
	\ln^2(\lambda)+\left(\beta^2-2\beta\gamma\right)\lambda^3\ln(\lambda)
	+\left(\gamma^2-\displaystyle\beta\gamma-\alpha_1\right)\lambda^3+o\left(\lambda^3\right)
	\end{equation}	
	
%	The results we obtained should match with the ones established recently by Gao and Lee (\cite{GL}). 
%	It seems to us that they give explicit formulas up to order $4$ for the expansion of the lognormal implied volatility 
%	in the case $T\rightarrow 0$ but we are too lazy to compare...
		
	\subsection{Basic definitions}
	In a Black-Scholes world, the dynamic of a stock $(S_t)$ is given by:
	$$
	d S_t = \sigma_{LN} S_t d W_t,
	$$
	with initial value $S$ at $t = 0$. The so-called "lognormal volatility" $\sigma_{LN}$ is related to the price of a call
	${\rm BS}\left(S,K,T,\sigma\right)$ struck at $K$ with maturity $T$ by the Black-Scholes formula (See \cite{BS}):
	
	\ba
	\label{BS}
	{\rm BS}\left(S,K,T,\sigma\right) = S N\left(d_+\right) - K N\left(d_-\right)
	\ea
	with
	$$
	N(x) =\frac{1}{\sqrt{2 \pi}}\int_{-\infty}^{x}  \exp \left({-\frac{u^2}{2}}\right) du
	$$
	and
	$$
	d_{\pm}=\displaystyle\frac{\ln\left(\displaystyle\frac{S}{K}\right)\pm
	\displaystyle\frac{\sigma_{LN}^2 T}{2}}{\sigma_{LN} \sqrt T}
	$$
	To simplify matters, we have considered $r=0$. Otherwise, we would consider the forward price 
	$F_t=S_t {\rm e}^{rt}$ instead of the spot price $S_t$. Following Ropper-Rutkowski (\cite{RR}), we set:

	\begin{defi}\label{def}
	Let us denote by:
	\begin{itemize}
	\item ${\rm TV} (S,K,K,T )$ (or simply ${\rm TV(T,K)}$ or ${\rm TV}$) the time-value of a European call option 
	struck at strike $K$ with maturity   $T$: ${\rm TV}\left(S,K,T,\sigma\right):={\rm BS}\left(S,K,T,\sigma\right)-(S-K)_+$ 
	\item $x := \ln (\frac{K}{S})$ (the log-moneyness)
	\item $\theta := \sigma_{LN}\sqrt{T}$ (the square root of the time-variance)
	\end{itemize}
	\end{defi}
	
	The spot price $S$ is assumed to be fixed by the market. 
	We will consider the two following cases: $K$ is fixed and $\sigma\sqrt{T}$ is small (case $1$) and
	$\sigma\sqrt{T}$ is fixed and $\displaystyle\frac{K}{S}$ is large (case $2$). In both cases, we will obtain a similar 
	expression for the asymptotic expansion of the implied lognormal volatility. 
	
	\section{Asymptotic expansions of a European call option}
	First let us assume that $x\not= 0$. 
	\subsection{Asymptotic expansions of a European call option for $x\not= 0$.}
	We note that the expression giving the time-value of a call-option in the case 
	($\theta\ll 1$ and $x$ fixed) is very similar to the case ($|x|\gg 1$ and $\theta$ fixed).
	
	\begin{prop}\label{prop}
	(Case $1$.) Let $N\in{\mathbb N}$. When $\theta\rightarrow 0$ and $x$ fixed, the asymptotic expansion of the time-value 
	$TV=C(T,K)-(S-K)_+$ of a call price 
	is given at order $N$ by:
	\begin{equation}
	\label{asymtv}
	4 \sqrt{\pi}\, \displaystyle\frac{{\rm e}^{-\frac{x}{2}}}{|x|}\, \left(\displaystyle\frac{TV}{S}\right) 
	=\left(\displaystyle\frac{2\theta^2}{x^2}\right)^{\frac{3}{2}}
	{\rm e}^{-\frac{1}{\left(\frac{2\theta^2}{x^2}\right)}}\,
	\displaystyle\sum_{k=0}^{N}\displaystyle\frac{(-1)^k}{2^k}\,
	a_k\left(\displaystyle\frac{x^2}{8}\right)\left(\displaystyle\frac{2\theta^2}{x^2}\right)^{k}
	+O\left(\theta^{2N+5}\,{\rm e}^{-\frac{x^2}{2\theta^2}}\right)
	\end{equation}
	
	with
	\begin{eqnarray}
	a_k(z)&:=&(2k+1)!!\,f_k(z)\\
	f_k(z)&:=&\displaystyle\sum_{j=0}^{k}\displaystyle\frac{z^j}{j!\,(2j+1)!!}
	\end{eqnarray}
	and for $j\in{\mathbb Z},\, (2j+1)!!:=\displaystyle\prod_{l=1}^{j}(2l+1)$ 
	(with the convention $\displaystyle\prod_{\emptyset}:=1$).
	
	\medskip
	\noindent
	(Case $2$.) Let $N\in{\mathbb N}$. When $|x|\rightarrow +\infty$ (i.e., $K\rightarrow 0$ or $K\rightarrow +\infty$) and
	$\theta$ fixed, the asymptotic expansion of the time-value of a call price 
	is given at order $N$ by:
	\begin{equation}
	\label{asymtvk}
	2\sqrt{2\pi}\, \displaystyle\frac{{\rm e}^{-\frac{x}{2}}}{\theta}\, \left(\displaystyle\frac{TV}{S}\right) = 
	\left(\displaystyle\frac{2\theta^2}{x^2}\right)\,{\rm e}^{-\frac{1}{\left(\frac{2\theta^2}{x^2}\right)}}
	\displaystyle\sum_{k=0}^{N}\, \displaystyle\frac{(-1)^k}{2^k}\,\, b_k\left(\displaystyle\frac{\theta^2}{4}\right)\,
	\left(\displaystyle\frac{2\theta^2}{x^2}\right)^{2k}\,
	+O\left(x^{-2N-4}\,{\rm e}^{-\frac{x^2}{2\theta^2}}\right)
	\end{equation}
	with
	\begin{equation}
	\label{defbk} 
	b_k(z) :=(2k+1)!!\displaystyle\sum_{j=0}^{k}\,(-1)^{j}\,
	{k\choose j}\,
	\displaystyle\frac{z^j}{(2j+1)!!}\,
	\end{equation}.	
	\bigskip

	\medskip
	\noindent
	(Case $3$.) Let $N\in{\mathbb N}$. When $\theta\rightarrow +\infty$ and $x$ fixed, 
	the asymptotic expansion of the covered call $CC=S-C(T,K)$ of a call price 
	is given at order $N$ by:
  \begin{eqnarray}
	\label{asymcc}	
  \sqrt{\pi}\, {\rm e}^{-\frac{x}{2}}\, \displaystyle\frac{{\rm CC}}{S}
  &=&\left(\displaystyle\frac{8}{\theta^2}\right)^{\frac{1}{2}}\,{\rm e}^{-\frac{1}{\left(\frac{8}{\theta^2}\right)}}
	\displaystyle\sum_{k=0}^{N}\,\displaystyle\frac{(-1)^k}{2^k}\,\,
	c_k\left(\displaystyle\frac{x^2}{8}\right)\,\left(\displaystyle\frac{8}{\theta^2}\right)^{k}
	+O\left(\theta^{-2N-3}\, {\rm e}^{-\frac{\theta^2}{8}}\right)  
  \end{eqnarray}
  with
	\begin{eqnarray}
	c_k(z)&:=&(2k-1)!!\,g_k(z)\\
	g_k(z)&:=&\displaystyle\sum_{j=0}^{k}\displaystyle\frac{z^j}{j!\,(2j-1)!!}
	\end{eqnarray}
	\end{prop}
	
	\noindent Note that $g'_k(z)=f_{k-1}(z)$ and $c'_k(z)=a_{k-1}(z)$.
	
	\bigskip
	\noindent {\bf Proof.} {\it Case $1$.}
 	For $n\in \mathbb{N}$, we denote by $\tilde{e}_n$ the function defined by 
	
	\begin{equation}\label{expoentilde}
	\forall\, u\in \mathbb{R},
	\quad {\rm e}^{-u}=1-u+\displaystyle\frac{u^2}{2!}-...+(-1)^n \displaystyle\frac{u^n}{n!}+{\tilde{e}}_n(u)
	\end{equation}
	
	Then, it is classical (properties of alternate series) that 
	
	\begin{equation}
	\label{class}
	\forall\, u>0,\quad |\tilde{e}_n(u)|\leq \displaystyle\frac{u^{n+1}}{(n+1)!}.
	\end{equation}

	Now, let us fix $N\in \mathbb{N}$. We start from:
	
  \begin{equation}
  \sqrt{2\pi}\, {\rm e}^{-\frac{x}{2}}\, \displaystyle\frac{TV}{S}=\displaystyle\int_0^{\theta} 
	{\rm e}^{-\frac{1}{2}\left(\frac{x^2}{{\xi}^2}+\frac{{\xi}^2}{4}\right)} d{\xi}
  \end{equation}
  with $x=\ln\left(\displaystyle\frac{K}{S}\right)$ as before. 
  This formula can be obtained by deriving the Black-Scholes formula with respect to $\theta$ 
  and then integrating the result (See [RR], Lemma 3.1). We have:
	
  \begin{equation}
  \label{rreq}
  \sqrt{2\pi}\, {\rm e}^{-\frac{x}{2}}\, \displaystyle\frac{TV}{S}
  =\displaystyle\int_0^{\theta}{\rm e}^{-\frac{x^2}{2 {\xi}^2}} {\rm e}^{-\frac{{\xi}^2}{8}} d{\xi}
  \end{equation}
	
	So, by (\ref{expoentilde}),
	
	$$
	\sqrt{2\pi}\, {\rm e}^{-\frac{x}{2}}\, \displaystyle\frac{TV}{S}
  =\displaystyle\sum_{n=0}^{N}\displaystyle\frac{(-1)^n}{n!\, 8^n}\,
  \displaystyle\int_0^{\theta}{\rm e}^{-\frac{x^2}{2 {\xi}^2}}\, {\xi}^{2n} d{\xi}
  +\displaystyle\int_0^{\theta}{\rm e}^{-\frac{x^2}{2 {\xi}^2}}\,\tilde{e}_N\left(\displaystyle\frac{{\xi}^2}{8}\right)\,d{\xi}
	$$
	
	So, with the change of variables $u:=\displaystyle\frac{x^2}{2 {\xi}^2}$, we get:
	
	\begin{eqnarray}
	\nonumber \sqrt{2\pi}\, {\rm e}^{-\frac{x}{2}}\, \displaystyle\frac{TV}{S}&=&
	\displaystyle\sum_{n=0}^{N}\displaystyle\frac{(-1)^n}{n!\, 16^n}\,\displaystyle\frac{|x|^{2n+1}}{2\,\sqrt{2}}
	\,\displaystyle\int_{\frac{x^2}{2\theta^2}}^{+\infty}\, u^{-n-\frac{3}{2}}{\rm e}^{-u}\, du + R_N(\theta)
	\\
	\label{devtvgammaseries}&=&
	\displaystyle\sum_{n=0}^{N}\displaystyle\frac{(-1)^n}{n!\, 16^n}\,\displaystyle\frac{|x|^{2n+1}}{2\,\sqrt{2}}
	\,\Gamma\left(-n-\displaystyle\frac{1}{2},\displaystyle\frac{x^2}{2\theta^2}\right) + R_N(\theta)
	\end{eqnarray}  

	with $R_N(\theta) := \displaystyle\int_{0}^{\theta}{\rm e}^{-\frac{x^2}{2 {\xi}^2}}\,
	\displaystyle\tilde{e}_n\left(\displaystyle\frac{{\xi}^2}{8}\right)\, d{\xi}$. 
	
	We have:

	\begin{eqnarray}
	\nonumber |R_N(\theta)|&\leq&
	\displaystyle\int_{0}^{\theta}{\rm e}^{-\frac{x^2}{2 {\xi}^2}}\,
	\displaystyle|\tilde{e}_N\left(\displaystyle\frac{{\xi}^2}{8}\right)|\, d{\xi}
	\\
	\nonumber	&\leq&\displaystyle\int_{0}^{\theta}{\rm e}^{-\frac{x^2}{2 {\xi}^2}}\,
	\left(\displaystyle\frac{1}{(N+1)!\, 8^{N+1}}\right) {\xi}^{2(N+1)} d{\xi}\\
	\nonumber &\leq&\displaystyle\frac{1}{(N+1)!\, 8^{N+1}}
	\displaystyle\int_{0}^{\theta}{\xi}^{2(N+1)}\, {\rm e}^{-\frac{x^2}{2 {\xi}^2}}\, d{\xi}\\
	\nonumber &\leq&\displaystyle\frac{1}{(N+1)!\, 8^{N+1}}
	\displaystyle\int_{\frac{x^2}{2 \theta^2}}^{+\infty} \left(\displaystyle\frac{x^2}{2}\right)^{N+1}\, 
	\displaystyle\frac{|x|}{2\sqrt{2}}\, u^{-(N+1)-\frac{3}{2}}\, {\rm e}^{-u}du \\
	\nonumber &\leq&\displaystyle\frac{|x|^{2N+3}}{2\sqrt{2}\, (N+1)!\,16^{N+1}}\,
	\Gamma\left(-N-\displaystyle\frac{3}{2},\displaystyle\frac{x^2}{2\theta^2}\right)
	\end{eqnarray} 	
	
	We recall the following asymptotic expansion valid for $z\rightarrow +\infty$ 
	and $m\in{\mathbb N}$	(see Formula 6.5.32 in (\cite{Abra})):
	
	\begin{eqnarray}
	\label{wellknown}\Gamma(a,z)&=&z^{a-1}\,{\rm e}^{-z}\displaystyle\left[1+\displaystyle\frac{a-1}{z}+...+
	\displaystyle\frac{(a-1)...(a-m)}{z^m}\right]+\gamma_m(a,z)
	\end{eqnarray} 		

	with	

	\begin{equation}\label{majogamma}
	\gamma_m(a,z) = O\left(\displaystyle\frac{z^{a-1}\,{\rm e}^{-z}}{z^{m+1}}\right)
	\end{equation}

	In particular, with $a=-N-\displaystyle\frac{3}{2},\, z=\displaystyle\frac{x^2}{2\theta^2}$ and $m=0$, 
	in the limit when $\theta\rightarrow 0$, we get:
	
	$$
	\Gamma\left(-N-\displaystyle\frac{3}{2},\displaystyle\frac{x^2}{2\theta^2}\right)=O\left[\left(
	\displaystyle\frac{x^2}{2\theta^2}\right)^{-N-\frac{5}{2}}\, {\rm e}^{-\frac{x^2}{2\theta^2}}\right].
	$$
	
	So, when $\theta\rightarrow 0$,
	
	\begin{equation}\label{majoRN}
	R_N(\theta) = O\left(\theta^{2N+5}\,{\rm e}^{-\frac{x^2}{2\theta^2}}\right)
	\end{equation}
	
	Moreover, for any $n<N$, we have by (\ref{wellknown}) with $m=N-n,\, a=-n-\frac{1}{2},\, z=\frac{x^2}{2\theta^2}$:

	\begin{eqnarray}
	\nonumber 	\Gamma\left(-n-\frac{1}{2},\frac{x^2}{2\theta^2}\right)&=&
	\left(\displaystyle\frac{x^2}{2\theta^2}\right)^{-n-\frac{3}{2}}{\rm e}^{-\frac{x^2}{2\theta^2}}\times\\
	\nonumber 	&\times&\left[1+\displaystyle\frac{\left(-n-\frac{1}{2}-1\right)}{\left(\frac{x^2}{2\theta^2}\right)^1}
	+...+ \displaystyle\frac{\left(-n-\frac{1}{2}-1\right)...\left(-n-\frac{1}{2}-(N-n)\right)}
	{\left(\frac{x^2}{2\theta^2}\right)^{N-n}}
	\right]+\\
	\nonumber	&+&\gamma_{N-n}\left(-n-\frac{1}{2},\frac{x^2}{2\theta^2}\right)\\
	\nonumber	&=& \left(\displaystyle\frac{2\theta^2}{x^2}\right)^{\frac{3}{2}}\times{\rm e}^{-\frac{x^2}{2\theta^2}}\\
	\nonumber	&\times&
	\left[\left(\displaystyle\frac{2\theta^2}{x^2}\right)^{n}+\displaystyle\frac{(-1)}{2}\displaystyle\frac{(2n+3)!!}{(2n+1)!!}
	\left(\displaystyle\frac{2\theta^2}{x^2}\right)^{n+1}+...+\displaystyle\frac{(-1)^{N-n}}{2^{N-n}}
	\displaystyle\frac{(2N+1)!!}{(2n+1)!!}
	\left(\displaystyle\frac{2\theta^2}{x^2}\right)^{N}
	\right]\\
	\nonumber &+&\gamma_{N-n}\left(-n-\frac{1}{2},\frac{x^2}{2\theta^2}\right)\\
	\nonumber &=&\left(\displaystyle\frac{2\theta^2}{x^2}\right)^{\frac{3}{2}}\times{\rm e}^{-\frac{x^2}{2\theta^2}}
	\left[\displaystyle\sum_{k=n}^{N}\displaystyle\frac{(-1)^{k-n}}{2^{k-n}}\displaystyle\frac{(2k+1)!!}{(2n+1)!!}
	\left(\displaystyle\frac{2\theta^2}{x^2}\right)^{k}	\right]\\
	\label{devgammaseries}&+&\gamma_{N-n}\left(-n-\frac{1}{2},\frac{x^2}{2\theta^2}\right) 	
	\end{eqnarray} 		
	
	Moreover, when $\theta\rightarrow 0$, we have by (\ref{majogamma}):
	
	\begin{eqnarray}
	\nonumber 	\gamma_{N-n}\left(-n-\frac{1}{2},\frac{x^2}{2\theta^2}\right)&=&
	O\left(\displaystyle\frac{\left(\displaystyle\frac{x^2}{2\theta^2}
	\right)^{-n-\frac{3}{2}}{\rm e}^{-\frac{x^2}{2\theta^2}}}
	{\left(\displaystyle\frac{x^2}{2\theta^2}\right)^{N-n+1}}\right)\\
	\label{ORN} &=&O\left(\theta^{2N+5}\,{\rm e}^{-\frac{x^2}{2\theta^2}}\right)
	\end{eqnarray}	
	
	Therefore, by (\ref{devtvgammaseries}), (\ref{majoRN}), (\ref{devgammaseries}), (\ref{ORN}), we obtain:
	
	\begin{eqnarray}
	\nonumber \sqrt{2\pi}\, \displaystyle\frac{{\rm e}^{-\frac{x}{2}}}{|x|}\, \displaystyle\frac{TV}{S} &=&
	\left(\displaystyle\frac{2\theta^2}{x^2}\right)^{\frac{3}{2}}{\rm e}^{-\frac{x^2}{2\theta^2}}\,
	\displaystyle\sum_{n=0}^{N}\displaystyle\frac{(-1)^n}{n!\,16^n}\,\displaystyle\frac{|x|^{2n}}{2\sqrt{2}}
	\displaystyle\sum_{k=n}^{N}\displaystyle\frac{(-1)^{k-n}}{2^{k-n}}\displaystyle\frac{(2k+1)!!}{(2n+1)!!}
	\left(\displaystyle\frac{2\theta^2}{x^2}\right)^{k}\\
	\nonumber &+&O\left(\theta^{2N+5}\,{\rm e}^{-\frac{x^2}{2\theta^2}}\right)
	\end{eqnarray}
	Hence,	
	\begin{equation}
	\nonumber 4 \sqrt{\pi}\, \displaystyle\frac{{\rm e}^{-\frac{x}{2}}}{|x|}\, \left(\displaystyle\frac{TV}{S}\right) 
	=\left(\displaystyle\frac{2\theta^2}{x^2}\right)^{\frac{3}{2}}{\rm e}^{-\frac{x^2}{2\theta^2}}\,
	\displaystyle\sum_{k=0}^{N}\,\displaystyle\frac{(-1)^k}{2^k}\,
	a_k\left(\displaystyle\frac{x^2}{8}\right)\,\left(\displaystyle\frac{2\theta^2}{x^2}\right)^{k}
	+O\left(\theta^{2N+5}\,{\rm e}^{-\frac{x^2}{2\theta^2}}\right)
	\end{equation}
	
	with
	\begin{equation}
	\nonumber a_k\left(\displaystyle\frac{x^2}{8}\right) := (2k+1)!!
	\displaystyle\sum_{j=0}^{k}\displaystyle\frac{1}{j!\,(2j+1)!!}\,\left(\displaystyle\frac{x^2}{8}\right)^j
	\end{equation}	
	which is exactly Proposition \ref{prop} - (\ref{asymtv}).
	
	\bigskip
	\noindent {\it Case $2$}: $\theta$ is fixed and $|x|\rightarrow +\infty$. Set:
	\begin{equation}
	\label{defix}
	I(x):=\int_{0}^{\theta}{\rm e}^{-\frac{x^2}{2\xi^2}}\, {\rm e}^{-\frac{\xi^2}{8}} d\xi. 
	\end{equation}
	With the help of the change of variables $z=\displaystyle\frac{\theta^2}{\xi^2}$, we have: 
	\begin{equation}
	\nonumber
	I(x)=\displaystyle\frac{\theta}{2}\,\displaystyle\int_{1}^{+\infty}\,{\rm e}^{-\frac{x^2}{2\theta^2} z}
	\,{\tilde h}(z)\,dz
	\end{equation}
	with ${\tilde h}(z):=\displaystyle\frac{16\sqrt{2}}{\theta^3}\,f_{-\frac{3}{2}}
	\left(\displaystyle\frac{8z}{\theta^2}\right)$ and $f_{\alpha}(z):=z^{\alpha}\,{\rm e}^{-\frac{1}{z}}$. 
	By induction on $n$, we show that 
	\begin{equation}
	\label{derivative}
	\forall n\in{\mathbb N},\, \forall z\in{\mathbb R}_+^*,\quad f_{\alpha}^{(n)}(z)=
	z^{\alpha-2n}{\rm e}^{-\frac{1}{z}}\displaystyle\sum_{p=0}^{n}
	{n\choose p} [\alpha-n+p]_p\,z^p
	\end{equation}
	where $f_{\alpha}^{(n)}$ is the $n^{th}$ derivative of $f_{\alpha}$ and with by definition, 
	$[u]_k:=\prod_{j=0}^{k-1}(u-j)$ 
	for any real $u$ and integer $k$.
	In particular, for any $(\alpha,N)\in{\mathbb R}_{-}\times {\mathbb N}^*$ fixed, 
	$f_{\alpha}^{(N)}(z)$ is uniformly bounded in $z\in{\mathbb R}_+^*$. Therefore, ${\tilde h}^{(N)}(z)$  is also 
	uniformly bounded in $z\in{\mathbb R}_+^*$. So $h^{(N)}(z)$ is also uniformly bounded in $z\in{\mathbb R}_+^*$ with 
	$h(z):={\tilde h}(z+1)$ (the function $h$ is analytic on ${\mathbb R}_+$), i.e.,
	\begin{equation}
	\label{derbounded}
	\forall N\in{\mathbb N}\, \exists\, M_N\in{\mathbb R}_{+}\, \forall z\in{\mathbb R}_{+},\quad h^{(N)}(z)\leq M_N
	\end{equation}	
	Let us fix $N\in{\mathbb N}$. By Taylor-Lagrange, we get: 
	\begin{eqnarray}
	\nonumber
	I(x)&=&\displaystyle\frac{\theta}{2}\,\displaystyle\int_{0}^{+\infty}\,{\rm e}^{-\frac{x^2}{2\theta^2}\, (z+1)}
	h(z)\,dz\\
	\nonumber
	&=&\displaystyle\frac{\theta}{2}\,
	{\rm e}^{-\frac{x^2}{2\theta^2}}\,\displaystyle\int_{0}^{+\infty}\,{\rm e}^{-\frac{x^2}{2\theta^2}\,z}
	\left(\displaystyle\sum_{k=0}^N \displaystyle\frac{h^{(k)}(0)}{k!}\, z^k+R_{N+1}(z)\right)\, dz
	\end{eqnarray}
	with $R_{N+1}(z)\leq M_{N+1} \displaystyle\frac{z^{N+1}}{(N+1)!}$. So,
	
	\begin{eqnarray}
  \nonumber
	I(x)&=&\displaystyle\frac{\theta}{2}\,
	{\rm e}^{-\frac{x^2}{2\theta^2}}\,\displaystyle\sum_{k=0}^{N}\,
	\displaystyle\frac{h^{(k)}(0)}{k!}\,\displaystyle\int_{0}^{+\infty}\,
	{\rm e}^{-\frac{x^2}{2\theta^2}\,z}\,z^k\, dz
	+ \displaystyle\frac{\theta}{2}\,
	{\rm e}^{-\frac{x^2}{2\theta^2}}\,\displaystyle\int_{0}^{+\infty}\,{\rm e}^{-\frac{x^2}{2\theta^2}\,z}\,R_{N+1}(z)\, dz
	\end{eqnarray}
	
	Using the fact that 
	\begin{equation}
	\forall A\in{\mathbb R}_{+}^{*},\quad
	\displaystyle\int_{0}^{+\infty}{\rm e}^{-A z} z^n\, dz = \displaystyle
	\frac{n!}{A^{n+1}}
	\end{equation}		
	we get:
	
	\begin{eqnarray}	
	\nonumber
	I(x)&=&\displaystyle\frac{\theta}{2}\,
	{\rm e}^{-\frac{x^2}{2\theta^2}}\,\displaystyle\sum_{k=0}^{N}\,\left[\displaystyle\frac{h^{(k)}(0)}{k!}\,
	k!\,\left(\displaystyle\frac{2\theta^2}{x^2}\right)^{k+1}	
	\right]+O\left({\rm e}^{-\frac{x^2}{2\theta^2}}\left(\displaystyle\frac{\theta^2}{x^2}\right)^{N+2}\right)\\
	\nonumber
	&=&\displaystyle\frac{\theta}{2}\,\left(\displaystyle\frac{2\theta^2}{x^2}\right)\,
	{\rm e}^{-\frac{x^2}{2\theta^2}}\,\displaystyle\sum_{k=0}^{N}\,{\tilde h}^{(k)}(1)\,
	\left(\displaystyle\frac{2\theta^2}{x^2}\right)^{k}
	+O\left({\rm e}^{-\frac{x^2}{2\theta^2}}\left(\displaystyle\frac{\theta^2}{x^2}\right)^{N+2}\right)\\
	\label{ix}
	&=&\displaystyle\frac{\theta}{2}\,\left(\displaystyle\frac{2\theta^2}{x^2}\right)\,
	{\rm e}^{-\frac{x^2}{2\theta^2}}\,\displaystyle\sum_{k=0}^{N}\,{\tilde h}^{(k)}(1)\,
	\left(\displaystyle\frac{2\theta^2}{x^2}\right)^{k}
	+O\left(x^{-2N-4}\,{\rm e}^{-\frac{x^2}{2\theta^2}}\right)
	\end{eqnarray}
	with 
	\begin{eqnarray}	
	\nonumber
	{\tilde h}^{(k)}(1)&:=&\displaystyle\frac{16\sqrt{2}}{\theta^3}\,
	\left(\displaystyle\frac{d^k}{z^k}\left[f_{-\frac{3}{2}}\,\displaystyle\frac{8z}{\theta^2}\right]\right)_{z=1}\\
	\nonumber
	&=&\displaystyle\frac{16\sqrt{2}}{\theta^3}\,\left(\displaystyle\frac{8}{\theta^2}\right)^{k}
	f_{-\frac{3}{2}}^{(k)}\,\left(\displaystyle\frac{8}{\theta^2}\right)\\
	\nonumber
	&=&\displaystyle\frac{16\sqrt{2}}{\theta^3}\,
	\left(\displaystyle\frac{8}{\theta^2}\right)^k\,
	\left(\displaystyle\frac{8}{\theta^2}\right)^{-\frac{3}{2}-2k}\,
	\displaystyle\sum_{j=0}^{k} {k\choose j}\,\left[-\frac{3}{2}-k+j\right]_j\,
	\left(\displaystyle\frac{8}{\theta^2}\right)^{j}\\
	\nonumber
	&=&\left(\displaystyle\frac{\theta^2}{8}\right)^{k}
	\displaystyle\sum_{j=0}^{k} {k\choose j}\,\left[-\frac{3}{2}-k+j\right]_j\,
	\left(\displaystyle\frac{\theta^2}{8}\right)^{-j}\\
	\nonumber
	&=&\displaystyle\sum_{j=0}^{k} {k\choose j}\,\left[-\frac{3}{2}-(k-j)\right]_j\,
	\left(\displaystyle\frac{\theta^2}{8}\right)^{k-j}\\
	\nonumber
	&=&\displaystyle\sum_{j=0}^{k} {k\choose j}\,\left[-\frac{3}{2}-j\right]_{k-j}\,
	\left(\displaystyle\frac{\theta^2}{8}\right)^{j}\\
	\nonumber
	&=&\displaystyle\sum_{j=0}^{k} {k\choose j}\,
	\displaystyle\frac{(-1)^{k-j}}{2^{k-j}}
	\displaystyle\frac{(2k+1)!!}{(2j+1)!!}\,
	\left(\displaystyle\frac{\theta^2}{8}\right)^{j}\\
	\label{devhtilde1}
	&=&\displaystyle\frac{(-1)^{k}}{2^{k}}
	\displaystyle\sum_{j=0}^{k}\,(-1)^{j}\,
	{k\choose j}\,
	\displaystyle\frac{(2k+1)!!}{(2j+1)!!}\,
	\left(\displaystyle\frac{\theta^2}{4}\right)^{j}
	\end{eqnarray}	
	Using (\ref{rreq}), (\ref{defix}), (\ref{ix}) and (\ref{devhtilde1}), this concludes the proof of Case $2$. 
	
	\bigskip
	\noindent
	{\it Case 3.}	
	We know turn to the case $\theta\rightarrow +\infty$.
	We start again from 
	\begin{equation}
  \nonumber
  \sqrt{2\pi}\, {\rm e}^{-\frac{x}{2}}\, \displaystyle\frac{C(T,K)-(S-K)_+}{S}=\displaystyle\int_0^{\theta} 
	{\rm e}^{-\frac{1}{2}\left(\frac{x^2}{{\xi}^2}+\frac{{\xi}^2}{4}\right)} d{\xi}
  \end{equation}
  
  When $\theta\rightarrow +\infty,\,C(T,K)\rightarrow S$. So,
  
  \begin{equation}
  \nonumber
  \sqrt{2\pi}\, {\rm e}^{-\frac{x}{2}}\, \displaystyle\frac{{\rm CC}}{S}=\displaystyle\int_{\theta}^{+\infty} 
	{\rm e}^{-\frac{1}{2}\left(\frac{x^2}{{\xi}^2}+\frac{{\xi}^2}{4}\right)} d{\xi}
  \end{equation}  
  
  with ${\rm CC}:= S-C(T,K)$. By the change of variables $\eta:=\displaystyle\frac{\xi^2}{8}$, we get:
  \begin{equation}
  \nonumber
  \sqrt{2\pi}\, {\rm e}^{-\frac{x}{2}}\, \displaystyle\frac{{\rm CC}}{S}=\displaystyle\int_{\frac{\theta^2}{8}}^{+\infty} 
	{\rm e}^{-\frac{x^2}{16\,\eta}}\,{\rm e}^{-\eta} \left(\sqrt{2}\eta^{-\frac{1}{2}}d{\eta}\right)
  \end{equation}   
  
  So, with the notations of (\ref{expoentilde}) and with $N\in{\mathbb N}^*$,
  \begin{eqnarray}
	\nonumber
  \sqrt{\pi}\, {\rm e}^{-\frac{x}{2}}\, \displaystyle\frac{{\rm CC}}{S}&=&
  \displaystyle\int_{\frac{\theta^2}{8}}^{+\infty}{\rm e}^{-\frac{x^2}{16}}\,\eta^{-\frac{1}{2}}\,{\rm e}^{-\eta}{d\eta}\\
	\nonumber  
  &=&\displaystyle\int_{\frac{\theta^2}{8}}^{+\infty}\left[\displaystyle\sum_{k=0}^{N}\displaystyle\frac{(-1)^k}{k!}
  \left(\displaystyle\frac{x^2}{16\eta}\right)^k+
  {\tilde e}_N\left(\displaystyle\frac{x^2}{16\,\eta}\right)\right]\eta^{-\frac{1}{2}}\,{\rm e}^{-\eta}d\eta\\
%	\nonumber
%  &=&\displaystyle\sum_{k=0}^{N}\displaystyle\frac{(-1)^k}{k!}
%  \displaystyle\int_{\frac{\theta^2}{8}}^{+\infty}\left(\displaystyle\frac{x^2}{16\eta}\right)^k
%  \,\eta^{-\frac{1}{2}}\,{\rm e}^{-\eta}d\eta
%  +\displaystyle\int_{\frac{\theta^2}{8}}^{+\infty}{\tilde e}_N
%  \left(\displaystyle\frac{x^2}{16\,\eta}\right)\eta^{-\frac{1}{2}}\,{\rm e}^{-\eta}d\eta\\
	\nonumber
  &=&\displaystyle\sum_{k=0}^{N}\displaystyle\frac{(-1)^k}{k!}
	\left(\displaystyle\frac{x^2}{16}\right)^k
  \displaystyle\int_{\frac{\theta^2}{8}}^{+\infty}\,\eta^{-k-\frac{1}{2}}\,{\rm e}^{-\eta}d\eta
  +\displaystyle\int_{\frac{\theta^2}{8}}^{+\infty}{\tilde e}_N
  \left(\displaystyle\frac{x^2}{16\,\eta}\right)\eta^{-\frac{1}{2}}\,{\rm e}^{-\eta}d\eta\\
	\label{tvlargetheta}
  &=&\displaystyle\sum_{k=0}^{N}\displaystyle\frac{(-1)^k}{k!}
	\left(\displaystyle\frac{x^2}{16}\right)^k
	\Gamma\left(-k+\frac{1}{2},\frac{\theta^2}{8}\right)
  +\displaystyle\int_{\frac{\theta^2}{8}}^{+\infty}{\tilde e}_N
  \left(\displaystyle\frac{x^2}{16\,\eta}\right)\eta^{-\frac{1}{2}}\,{\rm e}^{-\eta}d\eta
  \end{eqnarray}
  
  Moreover, for $\theta\rightarrow +\infty$, we have by (\ref{class}) and (\ref{majogamma}):
  \begin{eqnarray}
	\nonumber  
	\left|\displaystyle\int_{\frac{\theta^2}{8}}^{+\infty}{\tilde e}_N
  \left(\displaystyle\frac{x^2}{16\,\eta}\right)\eta^{-\frac{1}{2}}\,{\rm e}^{-\eta}d\eta	
	\right|&\leq&\displaystyle\int_{\frac{\theta^2}{8}}^{+\infty}
	\displaystyle\frac{1}{(N+1)!}
  \left(\displaystyle\frac{x^2}{16\,\eta}\right)^{N+1}\eta^{-\frac{1}{2}}\,{\rm e}^{-\eta}d\eta\\
	\nonumber  
	&\leq&
	\displaystyle\frac{1}{(N+1)!}\,
  \left(\displaystyle\frac{x^2}{16}\right)^{N+1}
	\Gamma\left(-N-\frac{1}{2},\frac{\theta^2}{8}\right)\\
	\nonumber	
	&=&O\left(\left(\displaystyle\frac{\theta^2}{8}\right)^{-N-\frac{3}{2}}
	{\rm e}^{-\frac{\theta^2}{8}}\right)\\	
	&=&O\left(\theta^{-2N-3}\, {\rm e}^{-\frac{\theta^2}{8}}\right)  		
  \end{eqnarray} 
  On the other hand, with the notations of (\ref{wellknown}) and $N,k\in{\mathbb N}$ with $N\geq k$,
  \begin{eqnarray}
	\nonumber   
	\Gamma\left(-k+\frac{1}{2},\frac{\theta^2}{8}\right)&=&\left(\displaystyle\frac{\theta^2}{8}\right)^{-k-\frac{1}{2}}
	\, {\rm e}^{-\frac{\theta^2}{8}}\left[1+\displaystyle\frac{-k-\frac{1}{2}}{\frac{\theta^2}{8}}
	+...+\displaystyle\frac{(-k-\frac{1}{2})...(-k-\frac{1}{2}-(N-k)+1)}{\left(\frac{\theta^2}{8}\right)^{N-k}}\right]\\
	\nonumber 
	&+&\gamma_{N-k}\left(-k+\displaystyle\frac{1}{2},\displaystyle\frac{\theta^2}{8}\right)\\
	&=&\left(\displaystyle\frac{\theta^2}{8}\right)^{-\frac{1}{2}}\,{\rm e}^{-\frac{\theta^2}{8}}
	\left[\displaystyle\sum_{j=k}^{N}
	\displaystyle\frac{\prod_{l=1}^{j-k}(-k+\frac{1}{2}-l)}{\left(\frac{\theta^2}{8}\right)^{j}}\right]
	+\gamma_{N-k}\left(-k+\displaystyle\frac{1}{2},\displaystyle\frac{\theta^2}{8}\right)
  \end{eqnarray}
  with 
	\begin{eqnarray}
	\nonumber   
	\left|\gamma_{N-k}\left(-k+\displaystyle\frac{1}{2},\displaystyle\frac{\theta^2}{8}\right)\right|&\leq&
	\displaystyle\frac{(-k-\frac{1}{2})...(-k-\frac{1}{2}-(N-k))}{\left(\frac{\theta^2}{8}\right)^{N-k+1}}
	\,{\rm e}^{-\frac{\theta^2}{8}}\\
	&=&O\left(\theta^{-2N-3}\, {\rm e}^{-\frac{\theta^2}{8}}\right)
  \end{eqnarray}
  Therefore,
  \begin{eqnarray}
	\nonumber
  \sqrt{\pi}\, {\rm e}^{-\frac{x}{2}}\, \displaystyle\frac{CC}{S}&=&
  \displaystyle\sum_{k=0}^{N}\displaystyle\frac{(-1)^k}{k!}
	\left(\displaystyle\frac{x^2}{16}\right)^k
	\left(\displaystyle\frac{\theta^2}{8}\right)^{-\frac{1}{2}}\,{\rm e}^{-\frac{\theta^2}{8}}
	\left[\displaystyle\sum_{j=k}^{N}
	\displaystyle\frac{\prod_{l=1}^{j-k}(-k+\frac{1}{2}-l)}{\left(\frac{\theta^2}{8}\right)^{j}}\right]\\
	\nonumber 
	&+&O\left(\theta^{-2N-3}\, {\rm e}^{-\frac{\theta^2}{8}}\right)\\
	\nonumber 	
	&=&\left(\displaystyle\frac{8}{\theta^2}\right)^{\frac{1}{2}}
	\,{\rm e}^{-\frac{\theta^2}{8}}
  \displaystyle\sum_{k=0}^{N}\displaystyle\frac{(-1)^k}{k!}
	\left(\displaystyle\frac{x^2}{16}\right)^k
	\left[\displaystyle\sum_{j=k}^{N}
	\displaystyle\frac{\prod_{l=1}^{j-k}(-k+\frac{1}{2}-l)}{\left(\frac{\theta^2}{8}\right)^{j}}\right]\\
	\nonumber	
	&+&O\left(\theta^{-2N-3}\, {\rm e}^{-\frac{\theta^2}{8}}\right)\\	
	\nonumber 	
	&=&\left(\displaystyle\frac{8}{\theta^2}\right)^{\frac{1}{2}}
	\,{\rm e}^{-\frac{\theta^2}{8}}
	\displaystyle\sum_{j=0}^{N}\left(\displaystyle\frac{8}{\theta^2}\right)^{j}
	\displaystyle\sum_{k=0}^{j}\displaystyle\frac{(-1)^k}{k!}
	\left(\displaystyle\frac{x^2}{16}\right)^k\displaystyle\prod_{l=1}^{j-k}(-k+\frac{1}{2}-l)\\
	&+&O\left(\theta^{-2N-3}\, {\rm e}^{-\frac{\theta^2}{8}}\right)
  \end{eqnarray}
  We have
  \begin{equation*}
  \label{cho}
  \displaystyle\prod_{l=1}^{j-k}(-k+\frac{1}{2}-l)=
  \displaystyle\frac{(-1)^{j-k}}{2^{j-k}}\displaystyle\frac{(2j-1)!!}{(2k-1)!!}  
  \end{equation*}
  Hence,
  \begin{eqnarray}
	\nonumber	
  \sqrt{\pi}\, {\rm e}^{-\frac{x}{2}}\, \displaystyle\frac{{\rm CC}}{S}
  &=&\left(\displaystyle\frac{8}{\theta^2}\right)^{\frac{1}{2}}\,{\rm e}^{-\frac{1}{\left(\frac{8}{\theta^2}\right)}}
	\displaystyle\sum_{j=0}^{N}\left(\displaystyle\frac{8}{\theta^2}\right)^{j}
	\displaystyle\sum_{k=0}^{j}\displaystyle\frac{(-1)^k}{k!}
	\left(\displaystyle\frac{x^2}{16}\right)^k
	\displaystyle\frac{(-1)^{j-k}}{2^{j-k}}\displaystyle\frac{(2j-1)!!}{(2k-1)!!}\\
	&+&O\left(\theta^{-2N-3}\, {\rm e}^{-\frac{\theta^2}{8}}\right)\\
	&=&\left(\displaystyle\frac{8}{\theta^2}\right)^{\frac{1}{2}}\,{\rm e}^{-\frac{1}{\left(\frac{8}{\theta^2}\right)}}
	\displaystyle\sum_{k=0}^{N}\,\displaystyle\frac{(-1)^k}{2^k}\,
	\,c_k\left(\displaystyle\frac{x^2}{8}\right),\left(\displaystyle\frac{8}{\theta^2}\right)^{k}
	+O\left(\theta^{-2N-3}\, {\rm e}^{-\frac{\theta^2}{8}}\right)  
  \end{eqnarray}
  with
  \begin{equation}
  \nonumber
  c_k\left(\displaystyle\frac{x^2}{8}\right)=
  (2k-1)!!\,\displaystyle\sum_{j=0}^{k}\displaystyle\frac{1}{j!\,(2j-1)!!}\,
  \left(\displaystyle\frac{x^2}{8}\right)^{j}
  \end{equation}
	This is exactly (\ref{asymcc}) and it puts an end to the proof of Proposition \ref{prop}.	
	\subsection{Asymptotic expansions of a European call option for $x= 0$.}
	At the money, we have $(S-K)_+=0$. So, ${\rm TV}=C$ and by (\ref{rreq}):
	\begin{equation}
	\nonumber
	\sqrt{2\pi}\,\displaystyle\frac{C}{S}=\displaystyle\int_{0}^{\theta}
	{\rm e}^{-\frac{\xi^2}{8}}\,d\xi
	\end{equation}
	So, 
	\begin{prop}
	At the money,
	\begin{eqnarray}
	\label{erfatm}
	C&=&S\,{\rm erf}\left(\displaystyle\frac{\theta}{2\sqrt{2}}\right)
	\end{eqnarray}
	
	with ${\rm erf}(u):=\displaystyle\frac{2}{\sqrt{\pi}}\displaystyle\int_{0}^{u}\,
	{\rm e}^{-\zeta^2}\,d\zeta$. 
	\end{prop}
	In the same way, we have:
	
	\begin{eqnarray}
	\displaystyle\frac{CC}{S}&=&{\rm erfc}\left(\displaystyle\frac{\theta}{2\sqrt{2}}\right)
	\end{eqnarray}
	with ${\rm erfc}(u):=\displaystyle\frac{2}{\sqrt{\pi}}\displaystyle\int_{u}^{+\infty}\,
	{\rm e}^{-\zeta^2}\,d\zeta$.

	\begin{prop}
	Let $N\in{\mathbb N}$.
	\noindent{\it (Case $1$.)} For $\theta\rightarrow 0$ and $\theta\not=0$, we have:
	\begin{equation}
	\label{dattvatm}
	\sqrt{2\pi}\,\displaystyle\frac{C}{S} = \theta\,
	\displaystyle\sum_{k=0}^{N}\displaystyle\frac{(-1)^k}{2^k}\,.\displaystyle\frac{1}{(2k+1)\,k!}\,
	\left(\displaystyle\frac{\theta^2}{4}\right)^{k} + O\left(\theta^{2N+3}\right)
	\end{equation}
	
	\medskip
	\noindent{\it (Case $2$.)} For $\theta\rightarrow +\infty$, we have:
	\begin{equation}
	\label{cctlarge}
	\sqrt{\pi}\,\displaystyle\frac{CC}{S} = 
	\left(\displaystyle\frac{8}{\theta^2}\right)^{\frac{1}{2}}\,{\rm e}^{-\frac{\theta^2}{8}}\,
	\displaystyle\sum_{k=0}^{N}\displaystyle\frac{(-1)^k}{2^k}\,.(2k-1)!!\,
	\left(\displaystyle\frac{8}{\theta^2}\right)^k+O\left(\theta^{-2N-3}\,{\rm e}^{-\frac{\theta^2}{8}}\right)
	\end{equation}
	\end{prop}	
	{\bf Proof.} Formula (\ref{cctlarge}) comes from the well known asymptotic expansion of ${\rm erfc}(x)$
	for $x$ large:
	\begin{equation}
	{\rm erfc}(x)=\displaystyle\frac{{\rm e}^{-x^2}}{x\,\sqrt{\pi}}\displaystyle\sum_{k=0}^{N}
	\displaystyle\frac{(-1)^k}{2^k}\,\displaystyle\frac{(2k-1)!!}{x^{2k}}
	+O\left(\displaystyle\frac{{\rm e}^{-x^2}}{x^{2N+3}}\right)
	\end{equation}
	\begin{note}
	\label{agree}
	Equation (\ref{cctlarge}) agrees with Proposition \ref{prop} - Equation (\ref{asymcc}) in the limit when
	$x\rightarrow 0$.
	\end{note}
		
	\section{Asymptotic expansions of the implied lognormal volatility}
	{\it This section is intended for people like me who are not familiar with the notion of transseries. 
	Otherwise, all the results below are supposed to be a simple consequence of the fact that 
	$\left(\lambda,\,\ln(\lambda)\right)$ form a transbase (See \cite{VdH}, Theorem $5.12$).} 

	\bigskip
	\noindent We want now to express the time-variance $\theta^2 = \sigma_{LN}^2 T$ in terms of the time-value $TV$ 
	(resp. covered call $CC$) for $\theta\ll 1$ 
	(resp. $\theta\gg 1$). 
	
	\subsection{Asymptotic expansions of the implied lognormal volatility when $K\not= S$}
	Let us assume that $K\not= S$ i.e., $x\not= 0$.	We need to invert Equations (\ref{asymtv}) and (\ref{asymcc}). 
		
	\medskip
	\noindent Our main result will be seen as a consequence of the following note.
	\begin{note}
	\label{thenote}	
	Both Equations (\ref{asymtv}) and (\ref{asymcc}) are of the form:
	\begin{equation}
	\label{tot}  
	v^{\beta}{\rm e}^{-\frac{1}{v}}\,\left[\displaystyle\sum_{k=0}^{N}\, \alpha_k\, 
	v^k\,+O\left(v^{N+1}\right)\right] ={\rm e}^{\gamma}\,{\rm e}^{-\frac{1}{\lambda}}  
	\end{equation}		
	with:
	\begin{itemize}
	\item (Case $\theta\ll 1$) 
	$v=\displaystyle\frac{2\,\theta^2}{x^2},\,\beta=\displaystyle\frac{3}{2},\,
	\alpha_k=\displaystyle\frac{(-1)^k}{2^k}\,.\,a_k\left(\displaystyle\frac{x^2}{8}\right),\,
	\gamma=\ln\left(\displaystyle\frac{4\sqrt{\pi}{\rm e}^{-\frac{x}{2}}}{|x|}\right)$ and
	$\lambda=-\displaystyle\frac{1}{\ln(\frac{TV}{S})}$.
	\item (Case $\theta\gg 1$) 
	$v=\displaystyle\frac{8}{\theta^2},\,\beta=\displaystyle\frac{1}{2},\,
	\alpha_k=\displaystyle\frac{(-1)^k}{2^k}\,.\,c_k\left(\displaystyle\frac{x^2}{8}\right),\,
	\gamma=\ln\left(\sqrt{\pi}\,{\rm e}^{-\frac{x}{2}}\right)$ and
	$\lambda=-\displaystyle\frac{1}{\ln(\frac{CC}{S})}$.
	\end{itemize}
	\end{note}
	
	\noindent We are going to invert (\ref{tot}) and thus to obtain an asymptotic expansion of $v$ 
	in terms of $\lambda^{\alpha}\ln(\lambda)^{\beta}$. 
	
	\begin{lemm}\label{da}
	For any $(\alpha_k)\in {\mathbb R}^{\mathbb N}, \gamma\in {\mathbb R}$ and $N\in{\mathbb N}^*$, 
	the asymptotics expansion of (\ref{tot}) for $0<\lambda\ll 1$ and $\beta>0$ is given by:
	\begin{equation}		
	\label{dae}
	v=\lambda-\beta\lambda^2
	\ln\lambda+\gamma\lambda^2+\beta^2\lambda^3
	\ln^2(\lambda)+\left(\beta^2-2\beta\gamma\right)\lambda^3\ln(\lambda)
	+\left(\gamma^2-\displaystyle\beta\gamma-\alpha_1\right)\lambda^3+o\left(\lambda^3\right)
	\end{equation}
	\end{lemm}
	
	\noindent{\bf Proof.} 
	
	{\bf Order $\lambda$}
	\medskip
		
	We use the fact that if $f\sim g$ with $\lim f = 0^+$ or $+\infty$, then also $\ln f\sim\ln g$. Therefore, from
	\begin{equation}\label{Cte}
	v^{\beta}{\rm e}^{-\frac{1}{v}}\sim {\rm e}^{\gamma}\,{\rm e}^{-\frac{1}{\lambda}}
	\end{equation}
	and the fact that 
	$\lim\limits_{\lambda \to 0}{\rm e}^{\gamma}\,{\rm e}^{-\frac{1}{\lambda}} =0$, we get:
	$$
	\beta \ln v-\displaystyle\frac{1}{v}\sim \gamma-\displaystyle\frac{1}{\lambda}
	$$
	The function $g: x\mapsto \beta\ln(x)-\displaystyle\frac{1}{x}$ is non-decreasing and
	$\lim\limits_{x \to 0^+} g(x)=-\infty$. So, $\lim\limits_{\lambda \to 0}v=0^+$.
	Moreover, since $\lim\limits_{v \to 0} v = 0$ and $\lim\limits_{v \to 0}v\ln v = 0$, we get $v\sim\lambda$. 
	
	\medskip
	\noindent{\bf Order $\lambda^2\ln(\lambda)$}
	\medskip
	
	Let us define $w$ by $v=\lambda(1+w)$.
	Necessarily, $\lim w=0$. Let us also denote by $\varepsilon$ the function such that $\lim \varepsilon =0$ and
	$$
	v^{\beta}{\rm e}^{-\frac{1}{v}}(1+\varepsilon) = {\rm e}^{\gamma}\,{\rm e}^{-\frac{1}{\lambda}}
	$$
	We have:
	$$
	\beta\ln v-\displaystyle\frac{1}{v}+\ln(1+\varepsilon)=\gamma -\displaystyle\frac{1}{\lambda}
	$$
	So,
	$$
	\displaystyle\frac{1}{v}-\displaystyle\frac{1}{\lambda}=\beta\ln v
	+\ln(1+\varepsilon)-\gamma
	$$
	The right hand side of the last equality is clearly equivalent to $\beta\ln v$ when $\lambda$ 
	(and so also $v$) goes to $0$.
	So,
	$$
	-\displaystyle\frac{w}{v}\sim \beta\ln v
	$$
	Thus,
	$$
	w\sim -\beta v\ln v\sim -\beta\lambda\ln \lambda
	$$
	So, we have proved:
	\begin{equation}
	v=\lambda-\beta\lambda^2\ln \lambda+o\left(\lambda^2\ln \lambda\right)
	\end{equation}
	
	\medskip
	\noindent{\bf Order $\lambda^2$}
	\medskip
	
	By (\ref{Cte}), we have:
	\begin{equation}\label{lnC}
	\beta\ln v-\displaystyle\frac{1}{v}= \gamma-\displaystyle\frac{1}{\lambda}+o(1)
	\end{equation}
	Set
	\begin{equation}\label{defz}
	v=\lambda\left(1-\beta\lambda\ln(\lambda)+z\right)
	\end{equation}
	with $z=o(\lambda\ln(\lambda))$.
	Then, 
	\begin{equation}\label{lnlamb}
	\ln(v)=\ln(\lambda)+o(1)
	\end{equation}
	and
	\begin{eqnarray}
	\nonumber
	\displaystyle\frac{1}{v}&=&\displaystyle\frac{1}{\lambda}\left[1-\beta\lambda\ln(\lambda)+z\right]^{-1}\\
	\nonumber 
	&=&\displaystyle\frac{1}{\lambda}\left(1+\beta\lambda\ln(\lambda)-z
	+o(\lambda)\right)\\
	\label{1v}
	&=&\displaystyle\frac{1}{\lambda}+\beta\ln(\lambda)-\displaystyle\frac{z}{\lambda}+o(1)
	\end{eqnarray}
	Therefore, by (\ref{lnC}),(\ref{lnlamb}) and (\ref{1v}), we obtain:
	$$
	\displaystyle\frac{z}{\lambda}=\gamma+o(1)
	$$
	So,
	\begin{equation}\label{symz}
	z\sim \gamma\lambda
	\end{equation}
	We have proved:
	\begin{equation}\label{symz}
	v=\lambda-\beta\lambda^2\ln \lambda+\gamma\lambda^2+o\left(\lambda^2\right)
	\end{equation}		
	
	\medskip
	\noindent{\bf Order $\lambda^3\ln^2(\lambda)$}
	\medskip
	
	Let $\xi$ be defined by 
	\begin{equation}
	v = \lambda\left(1-\beta\lambda\ln(\lambda)+\gamma\lambda+\xi\right)
	\end{equation}
	Then, $\xi=o(\lambda)$ and
	\begin{eqnarray}
	\nonumber
	\ln(v)&=&\ln(\lambda)+\ln\left(1-\beta\lambda\ln(\lambda)+\gamma\lambda+\xi\right)\\
	\nonumber 
	&=&\ln(\lambda)+O\left(\lambda\ln(\lambda)\right)\\
	\label{lnxi} 
	&=&\ln(\lambda)+o\left(\lambda\ln^2(\lambda)\right)
	\end{eqnarray}
	On the other hand,
	\begin{eqnarray}
	\nonumber
	\displaystyle\frac{1}{v}&=&\displaystyle\frac{1}{\lambda}
	\left[1-\beta\lambda\ln(\lambda)+\gamma\lambda+\xi\right]^{-1}\\
	\nonumber 
	&=&\displaystyle\frac{1}{\lambda}
	\left[1+\beta\lambda\ln(\lambda)-\gamma\lambda-\xi
	+\beta^2\lambda^2\ln^2(\lambda)+o\left(\lambda^2\ln^2(\lambda)\right)\right]\\
	\label{1xi}
	&=&\displaystyle\frac{1}{\lambda}+\beta\ln(\lambda)-\gamma-\displaystyle\frac{\xi}{\lambda}
	+\beta^2\lambda\ln^2(\lambda)+o\left(\lambda\ln^2(\lambda)\right)
	\end{eqnarray}
	With $N=1$, Equation (\ref{tot}) says that
	\begin{equation}
	\label{totoo}
	\beta\ln(v)-\displaystyle\frac{1}{v}+\ln\left(1+\alpha_1 v+o(v)\right)=\gamma-
	\displaystyle\frac{1}{\lambda}
	\end{equation}
	We have
	\begin{eqnarray}
	\nonumber
	\ln\left(1+\alpha_1 v+o(v)\right)&\sim&v\\
	\nonumber 
	&\sim&\lambda\\	
	\nonumber 
	&=&o\left(\lambda\ln^2(\lambda)\right)
	\end{eqnarray}
		
	So, by (\ref{lnxi}) and (\ref{1xi}), we get:
	\begin{equation}
	\displaystyle\frac{\xi}{\lambda}-\beta^2\lambda\ln^2(\lambda)+o\left(\lambda\ln^2(\lambda)\right)=0
	\end{equation}
	Therefore,
	\begin{equation}
	v=\lambda-\beta\lambda^2\ln \lambda
	+\gamma
	\lambda^2+\beta^2\lambda^3\ln^2(\lambda)+o\left(\lambda^3\ln^2(\lambda)\right)
	\end{equation}	
	
	\medskip
	\noindent{\bf Order $\lambda^3\ln(\lambda)$}
	\medskip
	
	Set $\phi$ so that
	\begin{equation}
	v=\lambda\left(1-\beta\lambda\ln(\lambda)+\gamma\lambda
	+\beta^2\lambda^2\ln^2(\lambda)+\phi\right)
	\end{equation}	
	with $\phi=o\left(\lambda^2\ln^2(\lambda)\right)$. We have:
	\begin{equation}
	\ln(v)=\ln(\lambda)-\beta\lambda\ln(\lambda)+o\left(\lambda\ln(\lambda)\right)
	\end{equation}	
	and
	\begin{eqnarray}
	\nonumber	
	\displaystyle\frac{1}{v}&=&\displaystyle\frac{1}{\lambda}
	\left(
	1-\beta\lambda\ln(\lambda)+\gamma\lambda+\beta^2\lambda^2\ln^2(\lambda)+\phi
	\right)^{-1}\\
	\nonumber	
	&=&\displaystyle\frac{1}{\lambda}
	\left(1+\beta\lambda\ln(\lambda)-\gamma\lambda-\beta^2\lambda^2\ln^2(\lambda)-\phi
	+\beta^2\lambda^2\ln^2(\lambda)-2\beta\gamma\lambda^2\ln(\lambda)+o\left(\lambda^2\ln(\lambda)\right)	\right)\\
	&=&\displaystyle\frac{1}{\lambda}+\beta\ln(\lambda)-\gamma-\displaystyle\frac{\phi}{\lambda}
	-2\beta\gamma\lambda\ln(\lambda)+o\left(\lambda\ln(\lambda)\right).
	\end{eqnarray}	

	\noindent So,
	$$
	\beta\ln(v)-\displaystyle\frac{1}{v}=\gamma-\displaystyle\frac{1}{\lambda}
	-\beta^2\lambda\ln(\lambda)+2\beta\gamma\lambda\ln(\lambda)+\displaystyle\frac{\phi}{\lambda}
	+o\left(\lambda\ln(\lambda)\right)
	$$

	\noindent On the other hand, we have:
	\begin{eqnarray}	
	\nonumber
	\ln\left(1+\alpha_1 v+o(v)\right)&\sim&\alpha_1 v\\
	\nonumber
	&\sim&\alpha_1\lambda\\
	&=&o\left(\lambda\ln(\lambda)\right)
	\end{eqnarray}		

	\noindent Thus, by (\ref{totoo}), we deduce that
	$$
	\displaystyle\frac{\phi}{\lambda}=\beta^2\lambda\ln(\lambda)-2\beta\gamma\lambda\ln(\lambda)+
	o\left(\lambda\ln(\lambda)\right)
	$$
	and so,
	$$
	\phi\sim\left(\beta^2-2\beta\gamma\right)\lambda^2\ln(\lambda)
	$$
	Therefore,
	\begin{equation}
	v=\lambda-\beta\lambda^2\ln \lambda
	+\gamma
	\lambda^2+\beta^2\lambda^3\ln^2(\lambda)+\left(\beta^2
	-2\beta\gamma\right)\lambda^3\ln(\lambda)+o\left(\lambda^3\ln(\lambda)\right)
	\end{equation}		

	\medskip
	\noindent{\bf Order $\lambda^3$}
	\medskip
	
	Set $\psi=o\left(\lambda^2\ln(\lambda)\right)$ such that:
	\begin{equation}
	v=\lambda\left(1-\beta\lambda\ln(\lambda)+\gamma\lambda
	+\beta^2\lambda^2\ln^2(\lambda)+\left(\beta^2
	-2\beta\gamma\right)\lambda^2\ln(\lambda)+\psi\right).
	\end{equation}
	Then,
	\begin{eqnarray}
	\nonumber
	\ln(v)&=&\ln(\lambda)+\ln\left(1-\beta\lambda\ln(\lambda)+\gamma\lambda
	+\beta^2\lambda^2\ln^2(\lambda)+\left(\beta^2
	-2\beta\gamma\right)\lambda^2\ln(\lambda)+\psi\right)\\
	&=&\ln(\lambda)-\beta\lambda\ln(\lambda)+\gamma\lambda+o(\lambda)\\
	\end{eqnarray}	
	Also,
	\begin{eqnarray}
	\nonumber
	\displaystyle\frac{1}{v}&=&\displaystyle\frac{1}{\lambda}
	\left(1-\beta\lambda\ln(\lambda)+\gamma\lambda
	+\beta^2\lambda^2\ln^2(\lambda)+\left(\beta^2
	-2\beta\gamma\right)\lambda^2\ln(\lambda)+\psi\right)^{-1}\\
	\nonumber
	&=&\displaystyle\frac{1}{\lambda}\left(1+\beta\lambda\ln(\lambda)
	-\gamma\lambda-\beta^2\lambda^2\ln^2(\lambda)
	-\left(\beta^2-2\beta\gamma\right)\lambda^2\ln(\lambda)-\psi\right)\\
	\nonumber&+&
	\displaystyle\frac{1}{\lambda}\left(
	\beta^2\lambda^2\ln^2(\lambda)-2\beta\gamma\lambda^2\ln(\lambda)+\gamma^2\lambda^2+o(\lambda^2)\right)\\
	\nonumber
	&=&\displaystyle\frac{1}{\lambda}+\beta\ln(\lambda)-\gamma-\beta^2\lambda\ln(\lambda)+\gamma^2\lambda-
	\displaystyle\frac{\psi}{\lambda}+o(\lambda)
	\end{eqnarray}	
	and
	\begin{eqnarray}
	\nonumber
	\ln(1+\alpha_1 v+o(v))&=&\alpha_1 v+o(v)\\
	&=&\alpha_1 \lambda+o(\lambda)
	\end{eqnarray}
	Therefore,
	\begin{equation}
	\nonumber
	\beta\ln(v)-\displaystyle\frac{1}{v}+\ln(1+\alpha_1 v+o(v))-\gamma+\displaystyle\frac{1}{\lambda}=
	\beta\gamma\lambda-\gamma^2\lambda+\displaystyle\frac{\psi}{\lambda}+\alpha_1\lambda+o(\lambda)
	\end{equation}
	By (\ref{totoo}), the left hand side of this equation is $0$. So,
	\begin{equation}
	\nonumber
	\displaystyle\frac{\psi}{\lambda}=\left(\gamma^2-\beta\gamma-\alpha_1\right)\lambda+o(\lambda)
	\end{equation}
	and
	\begin{equation}
	\nonumber
	\psi=\left(\gamma^2-\beta\gamma-\alpha_1\right)\lambda^2+o(\lambda^2)
	\end{equation}
	This put an end to Lemma \ref{da}. 
	
	\medskip
	\noindent
	By induction on $m$ and $n$, we can also prove the following generalization 
	of Lemma \ref{da}.
	
	\begin{prop}
	There are $a_{i,j}$ defined for $(i,j)^2\in{\mathbb N}$ and $j<i$ such that for any $(m,n)\in{\mathbb N}^2$, with 
	$n<m$, we have:
	\begin{equation}
	v = v_{m,n}+o\left(\lambda^m\ln^n(\lambda)\right)
	\end{equation}
	with
	\begin{equation}
	v_{m,n} := \displaystyle\sum_{i=1}^{m}\sum_{j=n}^{m-1}a_{i,j}\,\lambda^i\ln^j(\lambda)
	\end{equation}
	\begin{itemize}
	\item We have $\lambda\succ\lambda^2\ln(\lambda)\succ\lambda^2\succ\lambda^3\ln^2(\lambda)
	\succ\lambda^3\ln(\lambda)\succ\lambda^3\succ\lambda^4\ln^3(\lambda)\succ...$. The symbol $\succ$ is defined by 
	$f\succ g$ if and only if $g=o(f)$ in a neighborhood of $0$.
	\item In this sequence, $\lambda^i\ln^j(\lambda)$ is in position $\pi_{i,j}:=\displaystyle\frac{i(i+1)}{2}-j$.
	\item for any $k\in{\mathbb N}$, there is a unique $(i,j)\in{\mathbb N}$ with $j<i$ such that $k=\pi_{i,j}$.
	\item we set $v_k:=v_{i,j}$ with $k=\pi_{i,j}$.
	\end{itemize}	
	
	\noindent
	For $m\geq 3$, the coeficient $a_{m,n}$ can be obtained by induction by the following way:
	\begin{itemize}
	\item We expand $\ln\left(\displaystyle\frac{v_{\pi_{m,n}-1}}{\lambda}\right),\, \displaystyle\frac{1}{v_{\pi_{m,n}-1}}$ and 
	$\ln\left(\displaystyle\sum_{k=0}^{m-2}\alpha_k v^k\right)$ and we keep the terms in $\lambda^{m-2}\ln^n(\lambda)$.
	\item We note those terms $A_{m,n}, B_{m,n}$ and $C_{m,n}$ respectively.
	\item Then, $a_{m,n}=B_{m,n}-\displaystyle\frac{3}{2}A_{m,n}-C_{m,n}$.
	\end{itemize}
	\end{prop}
	
	\noindent
	As an application of Note \ref{thenote} and Lemma \ref{da}, we get:
	\begin{prop}\label{laprop}
	{\it (Case $1$: short expiry)}. Let us denote by $TV:=C(T,K)-(S-K)_+$ the time-value of a European call option, 
	$\sigma_{LN}$ its implied lognormal volatility 
	and $T$ the maturity of the option. Set $\lambda:=-\displaystyle\frac{1}{\ln(\frac{TV}{S})},
	\,\gamma:=\ln\left(\displaystyle\frac{4\sqrt{\pi}{\rm e}^{-\frac{x}{2}}}{|x|}\right)$ and 
	$\alpha_1=-\displaystyle\frac{3}{2}-\displaystyle\frac{x^2}{16}$ with $x=\ln (\frac{K}{S})$.
	Then, when $T\rightarrow 0$, we have the following expansion for the time-variance of the call option: 
	$\sigma_{LN}^2 T=\displaystyle\frac{x^2}{2}\, v$ with
	$$
	v = \lambda-\displaystyle\frac{3}{2}\lambda^2\ln \lambda
	+\gamma\lambda^2+\displaystyle\frac{9}{4}\lambda^3\ln^2(\lambda)+
	\left(\displaystyle\frac{9}{4}
	-3\gamma\right)\lambda^3\ln(\lambda)+\left(\gamma^2-\displaystyle\frac{3}{2}\gamma-\alpha_1\right)\lambda^3
	+o\left(\lambda^3\right)
	$$
	{\it (Case $2$: large expiry)}. Let us denote by $CC:=S-C(T,K)$ the covered call of a European call option, 
	$\sigma_{LN}$ its implied lognormal volatility and $T$ the maturity of the option. 
	Set $\lambda:=-\displaystyle\frac{1}{\ln(\frac{CC}{S})},
	\,\gamma:=\ln\left(\sqrt{\pi}\,{\rm e}^{-\frac{x}{2}}\right)$ and 
	$\alpha_1=-\displaystyle\frac{1}{2}-\displaystyle\frac{x^2}{16}$ with $x=\ln (\frac{K}{S})$.
	Then, when $T\rightarrow +\infty$, we have the following expansion for the time-variance of the call option:
	\begin{equation}
	\label{geneT}
	\sigma_{LN}^{2}T=\displaystyle\frac{8}{\lambda}
	\left[1+\displaystyle\frac{1}{2}\lambda\ln(\lambda)-\gamma\lambda-\displaystyle\frac{1}{4}\lambda^2\ln(\lambda)
	+\left(\displaystyle\frac{\gamma}{2}+\alpha_1\right)\lambda^2+o\left(\lambda^2\right)\right]
	\end{equation}
	In particular, when $T\rightarrow +\infty,\quad \sigma_{LN}\sim 
	2\sqrt{-\displaystyle\frac{2\ln\left(\frac{CC}{S}\right)}{T}}$.
	\end{prop}
	
	\noindent 
	Equation (\ref{geneT}) is a generalization of \cite{T}.
	
	\medskip
	\noindent {\bf Proof.} Case $1$ is just an application of Note \ref{thenote} and Lemma \ref{da}.
	Case $2$ follows from the expansion of $v^{-1}$. Indeed, by (\ref{dae}), we have:
	\begin{equation}
	\displaystyle\frac{1}{v}=\displaystyle\frac{1}{\lambda}
	\left[1+\beta\lambda\ln(\lambda)-\gamma\lambda-\beta^2\lambda^2\ln(\lambda)+(\beta\gamma+\alpha_1)\lambda^2
	+o\left(\lambda^2\right)\right]
	\end{equation}
	Therefore using Note \ref{thenote} - Case $2$,  
	\begin{equation}
	\nonumber
	\displaystyle\frac{\sigma_{LN}^{2}T}{8}=\displaystyle\frac{1}{\lambda}
	\left[1+\displaystyle\frac{1}{2}\lambda\ln(\lambda)-\gamma\lambda-\displaystyle\frac{1}{4}\lambda^2\ln(\lambda)
	+(\displaystyle\frac{\gamma}{2}+\alpha_1)\lambda^2+o\left(\lambda^2\right)\right]
	\end{equation}
	Hence, we get the result.		
	
	\begin{note}
	The case $x\rightarrow +\infty$ and $\theta$ fixed (Case $2$ of Proposition \ref{prop}) 
	can be treated exactly in the same way. It is more or less exactly the same as the case $x$ fixed and 
	$\theta\rightarrow 0$ except that $\beta$ is now equal to $1$,
	$\gamma$ has to be replaced by $\ln\left(2\sqrt{2\pi}\, \displaystyle\frac{{\rm e}^{-\frac{x}{2}}}{\theta}\right)$,  
	and $\alpha_k$ ($k\in{\mathbb N}$) has to be replaced by 
	$\displaystyle\frac{(-1)^k}{2^k}\,\, b_k\left(\displaystyle\frac{\theta^2}{4}\right)$ with $b_k$ given in (\ref{defbk}). 
	Therefore, $\alpha_1=-\displaystyle\frac{3}{2}+\displaystyle\frac{x^2}{8}$ and the formula for the 
	implied lognormal volatility is $\sigma_{LN}^2=\displaystyle\frac{x^2}{2}\,v$ with
	\begin{equation}		
	\label{sigwlargeda}
	v=\lambda-\lambda^2\,\ln\lambda+\gamma\lambda^2+\lambda^3\,\ln^2(\lambda)
	+\left(1-2\gamma\right)\lambda^3\ln(\lambda)
	+\left(\gamma^2-\gamma-\alpha_1\right)\lambda^3+o\left(\lambda^3\right)
	\end{equation}
	\end{note}
	
	\subsection{Implied lognormal volatility at the money}
	It turns out that at the money, there is a closed form formula for implied lognormal volatility in terms of call price.
	No asumption on $T$ is made.
	
	\begin{prop}
	\label{daatm}
	At the money, implied lognormal volatility $\sigma_{LN}$ 
	can be obtained as a power series in call price $C$ according to the formula:
	
	\begin{equation}
	\sigma_{LN}=\sqrt{\displaystyle\frac{2\pi}{T}}\,\displaystyle\frac{C}{S}
	\displaystyle\sum_{k=0}^{\infty}
	\displaystyle\frac{\pi^k\eta_k}{4^k(2k+1)}
	\left(\displaystyle\frac{C}{S}\right)^{2k}
	\end{equation}
	with $\eta_k$ given by induction:
	\begin{equation}
	\eta_k=\displaystyle\sum_{j=0}^{k}\displaystyle\frac{\eta_j\,\eta_{k-1-j}}{(j+1)(2j+1)}
	\end{equation}
	\end{prop}
	
	\noindent
	{\bf Proof.} We have the well known expansion of ${\rm erf}^{-1}$ (see for instance \cite{Abra}):
	\begin{eqnarray}
	{\rm erf}^{-1}(x)&=&\displaystyle\sum_{k=0}^{\infty}
	\displaystyle\frac{\eta_k}{2k+1}
	\left(\displaystyle\frac{\sqrt{\pi}}{2}\,x\right)^{2k+1}\\
	\eta_k&=&\displaystyle\sum_{j=0}^{k}\displaystyle\frac{\eta_j\,\eta_{k-1-j}}{(j+1)(2j+1)}
	\end{eqnarray}
	So, by (\ref{erfatm}),
	\begin{eqnarray}
	\nonumber
	\theta&=&2\sqrt{2}\,\displaystyle\frac{\sqrt{\pi}}{2}\,\displaystyle\frac{C}{S}
	\displaystyle\sum_{k=0}^{\infty}
	\displaystyle\frac{\eta_k}{(2k+1)}
	\left(\displaystyle\frac{\sqrt{\pi}}{2}\right)^{2k}
	\left(\displaystyle\frac{C}{S}\right)^{2k}\\
	&=&\sqrt{2\pi}\,\displaystyle\frac{C}{S}
	\displaystyle\sum_{k=0}^{\infty}
	\displaystyle\frac{\pi^k\eta_k}{4^k(2k+1)}
	\left(\displaystyle\frac{C}{S}\right)^{2k}
	\end{eqnarray}


\begin{thebibliography}{}

\bibitem{Abra}
Abramowitz M., Stegun IA. (1966), ``Handbook of Mathematical Functions'', Dover. \smallskip

\bibitem{BS}  
F. Black and M. Scholes (1973), ``The Pricing of Options and Corporate Liabilities'', {\it J. Political Economy} 81:  637-654 \smallskip 

\bibitem{GL}
K. Gao, R. Lee, ``Asymptotics of Implied Volatility to Arbitrary Order'', ssrn 1768383. \smallskip

\bibitem{RR}
Roper M., Rutkowski M. (2009), ``A Note on the Behaviour of the Black-Scholes Implied Volatility Close to Expiry'', {\it Int. J. of Theoretical and Applied Finance}, 12(4): 427 –- 441.  \smallskip

\bibitem{T}
M. Tehranchi (2009), Asymptotics of Implied Volatility Far From Maturity. {\it J. of Applied Probability},
46(3):629-650. \smallskip

\bibitem{VdH}
Van der Hoeven, J. (2006), ``Transseries and Real Differential Algebra''. {\it Lecture Notes in Math.} 1888, Springer.

\end{thebibliography}
	\end{document}